\documentclass[12pt]{iopart}
\usepackage{wasysym,amssymb}

\newcommand{\be}{\begin{equation}}
\newcommand{\ee}{\end{equation}}
\newcommand{\bea}{\begin{eqnarray}}
\newcommand{\eea}{\end{eqnarray}}

\newcommand{\D}{\mathcal{D}}

\newcommand{\m}{\, \mathrm{m}}
\newcommand{\cm}{\, \mathrm{cm}}
\newcommand{\s}{\, \mathrm{s}}
\newcommand{\km}{\, \mathrm{km}}
\newcommand{\kg}{\, \mathrm{kg}}
\newcommand{\g}{\, \mathrm{g}}
\newcommand{\AU}{\, \mathrm{AU}}
\newcommand{\Mpc}{\, \mathrm{Mpc}}
\newcommand{\eV}{\, \mathrm{eV}}

\begin{document}

\title{Solar system constraints on \\ Gauss-Bonnet mediated dark energy}

\author{Luca Amendola$^1$, Christos Charmousis$^2$ and Stephen C Davis$^3$}

\address{${}^1$ INAF/Osservatorio Astronomico di Roma, 
Viale Frascati 33, 00040 Monte Porzio Catone (Roma), Italy}

\address{${}^2$ LPT, Universit\'{e} Paris--Sud, B\^{a}timent 210, 
91405 Orsay CEDEX, France}

\address{${}^3$ Lorentz Institute, Postbus 9506, 2300 RA Leiden, 
The Netherlands}

\eads{\mailto{amendola@mporzio.astro.it},
\mailto{Christos.Charmousis@th.u-psud.fr} and
\mailto{sdavis@lorentz.leidenuniv.nl}
 }

\begin{abstract}
Although the Gauss-Bonnet term is a topological invariant for general
relativity, it couples naturally to a quintessence scalar field,
modifying gravity at solar system scales. We determine the 
solar system constraints due to this term by evaluating the
post-Newtonian metric for a distributional source. We find a mass
dependent, $1/r^7$ correction to the Newtonian potential, and also 
deviations from the Einstein gravity prediction for light-bending. We
constrain the parameters of the theory using planetary orbits, the
Cassini spacecraft data,  and a laboratory test of Newton's law, always finding
extremely tight bounds on the energy associated to the Gauss-Bonnet term. We
discuss the relevance of these constraints to late-time cosmological
acceleration.

\end{abstract}

\section{Introduction}

Supernovae measurements~\cite{supernova} indicate that our universe
has entered a phase of late-time acceleration. One can question the
magnitude of the acceleration and its equation of state, although
given the concordance of different cosmological data, acceleration
seems a robust observation (although see~\cite{sarkar} for
criticisms). Commonly, in order to explain this phenomenon one
postulates the existence of a minute cosmological constant $\Lambda
\sim 10^{-12} \eV^4$. This fits the data well and is the most
economic explanation in terms of parameter(s). However such a tiny
value is extremely unnatural from a particle physics point of 
view~\cite{weinberg}. Given the theoretical problems of a cosmological
constant, one hopes that the intriguing phenomenon of acceleration is
a window to new observable physics. This could be in the matter
sector, in the form of dark energy~\cite{quint,de}, or in the gravity
sector, in the form of a large distance modification of
Einstein gravity~\cite{mg,dgp,deffayet,f(r)}.
Scalar field driven dark energy, or quintessence~\cite{quint} is one
of the most popular of the former possibilities. However these  models
have important drawbacks, such as the fine tuning of the mass of the
quintessence field (which has to be smaller than the actual Hubble
parameter, $H_0\sim 10^{-33} \eV$), and stability of radiative
corrections from the matter sector~\cite{lorenzo} 
(see however~\cite{lorenzo2}). Modified gravity models have the
potential to avoid 
these problems, and can give a more profound explanation of the
acceleration. However, these are far more difficult to obtain since
Einstein's theory is experimentally well established~\cite{will}, and
the required modifications happen at very low (classical) energy
scales which are (supposed to be) theoretically well
understood. Furthermore, many apparently successful modified gravity
models suffer from instabilities or are incompatible with gravity
experiments. For example the  self-accelerating solutions of
DGP~\cite{deffayet} suffer from perturbative ghosts~\cite{killdgp},
and $f(R)$ gravity theories~\cite{f(r)} can conflict with solar system
measurements and present instabilities~\cite{killfr}.

In this paper we will consider observational constraints on a class of
gravity theories which feature both dark energy and modified
gravity. Specifically, we will  examine solar system and laboratory
constraints resulting from the response of gravity to a
quintessence-like scalar field, which couples to quadratic order curvature
terms such as the Gauss-Bonnet term. Such couplings arise
naturally~\cite{us}, and modify gravity at local and cosmological
scales~\cite{us,mota}. Although the Gauss-Bonnet invariant shares many of the
properties of the Einstein-Hilbert term, the resulting theory can have
substantially different features, see for example~\cite{daviz}. It is
a promising candidate for a consistent explanation of cosmological
acceleration, but as we will show, can also produce undesirable
effects at solar system scales.

In particular, we will determine constraints from deviations in
planetary orbits around the sun, the frequency shift of signals from
the Cassini probe, and table-top experiments. In contrast to some previous
efforts in the field~\cite{GBPPN}, we will not suppose {\em a priori} the
order of the Gauss-Bonnet correction or the scalar field
potential. Instead we will calculate leading-order gravity corrections
for each of them, and obtain constraints on the relevant coupling constants
(checking they fall within the validity of our perturbative
expansion). Hence our analysis will apply for large couplings, which
as we will see, are in accord with Gauss-Bonnet driven effective dark
energy models. In this way we will show such models generally produce
significant deviations from general relativity at local scales. We
also include higher-order scalar field kinetic terms, although for the
solutions we consider they turn out to be subdominant.
 
In the next section we will present the theory in question and
calculate the corrections to a post-Newtonian metric for a distributional point
mass source. In section~\ref{sec:con}, we derive constraints 
from planetary motion, the Cassini probe, and a table-top
experiment. For the Cassini constraint, we have to explicitly derive
the predicted frequency shift for our theory, as it does not fall within the
usual Parametrised Post-Newtonian (PPN) analysis. We discuss the
implications of our results in section~\ref{sec:dis}.

\section{Quadratic Curvature Gravity }
\label{sec:grav}

We will consider a theory with the second-order gravitational Lagrangian
\bea \fl
\mathcal{L} = \sqrt{-g} \, \biggl\{R - (\nabla \phi)^2 -2V(\phi) 
\nonumber \\ \fl \hspace{0.8in} {}
+ \alpha\left[
\xi_1(\phi)  \mathcal{L}_{GB}
+ \xi_2(\phi) G^{\mu\nu}\nabla_{\! \mu} \phi \nabla_{\! \nu} \phi
+\xi_3(\phi)(\nabla \phi)^2\nabla ^2\phi +\xi_4(\phi)  (\nabla \phi)^4 \right]
 \biggr\}  \, ,
\label{Lag}
\eea
which includes the Gauss-Bonnet term $\mathcal{L}_{GB} = 
R^2 - 4R_{\mu\nu} R^{\mu\nu} + R_{\mu\nu\rho\sigma} R^{\mu\nu\rho\sigma}$.
Note for example that such a Lagrangian with given $\xi$'s arises
naturally from higher dimensional compactification of a pure gravitational
theory~\cite{us}. On its own, in four dimensions, the Gauss-Bonnet
term does not contribute to the gravitational field equations. However
we emphasise that when coupled to a scalar field (as above), it has a
non-trivial effect.

Throughout this paper we take the dimensionless couplings $\xi_i$ and
their derivatives to be $\Or(1)$. There is then only one scale for the
higher curvature part of the action, given by the parameter $\alpha$,
with dimensions of length squared. Similarly we assume that all
derivatives of the potential $V$ are of $\Or(V)$, which in our
conventions has dimensions of inverse length squared. These two
simplifying assumptions will hold for a wide range of theories,
including those in which $\xi_i$ and $V$ arise from a toroidal
compactification of a higher dimensional space~\cite{us}. On the other
hand it is perfectly conceivable that they do not apply for our
universe, in which case the corresponding gravity theories will not be
covered by the analysis in this article.

Using the post-Newtonian limit, the metric for the solar system can be
written~\cite{will}
\be
ds^2= -(1-h_{00})(c\, dt)^2  + (\delta_{ij}+h_{ij}) dx^i dx^j 
+ \Or(\epsilon^{3/2}) \, .
\ee
with $h_{00}, h_{ij} = \Or(\epsilon)$. The dimensionless parameter
$\epsilon$ is the typical gravitational strength, given by 
$\epsilon=G m/(r c^2)$ where $m$ is the typical mass scale and $r$ the typical
length scale (see below). For the solar system $\epsilon$ is at most
$10^{-5}$, while for cosmology, or close to the event horizon of a
black hole, it is of order unity. The scale of planetary
velocities $v$, is of order $\epsilon^{1/2}$, and so the
$h_{0i}$ components of the metric are $\Or(\epsilon^{3/2})$, as are
$\partial_t h_{00}$ and $\partial_t h_{ij}$. In what follows, we will
take $\phi =\phi_0+\Or(\epsilon)$. For the linearised approximation we
are using, we can adopt a post-Newtonian gauge in which the off-diagonal
components of $h_{ij}$ are zero. We can then write
\be
h_{ij} = -2 \Psi \delta_{ij} \, , \qquad  
h_{00} = -2 \Phi \, ,
\ee
and so $c^2 \Phi$ is the Newtonian potential.

In this paper we will consider the leading-order
corrections in $\epsilon$ without assumptions on the magnitude of $V$
and $\alpha$. To leading order in $\epsilon$, the Einstein
equations take the nice compact form,
\bea
\fl \Delta \Phi = \frac{4\pi G_0}{c^2} \rho_m - V
- 2\alpha  \xi_1'  \D(\Phi+\Psi,\phi) 
+ \Or(\epsilon^2,\alpha \epsilon^3/r^2, V \epsilon r^2)
\\
\fl \Delta \Psi = \frac{4\pi G_0}{c^2} \rho_m + \frac{V}{2}
- \alpha \left[ 2\xi_1'  \D(\Psi,\phi) 
+\frac{\xi_2}{4}\D(\phi,\phi) \right] 
+ \Or(\epsilon^2,\alpha \epsilon^3/r^2, V \epsilon r^2)
\eea
where primes denote $\partial/\partial \phi$, and $V$, $\xi_1'$, etc.\ are
evaluated at $\phi=\phi_0$. The matter energy density in
the solar system is $\rho_m$, and $G_0$ is its bare coupling strength (without
quadratic gravity corrections). Other components of the
energy-momentum tensor are higher order in $\epsilon$. The scalar
field equation is
\be
\fl \Delta \phi=V'  - \alpha \left[4\xi_1'  \D(\Phi,\Psi) 
+ \xi_2 \D(\Phi-\Psi,\phi) + \xi_3 \D(\phi,\phi)\right]
+ \Or(\epsilon^2,\alpha \epsilon^3/r^2, V \epsilon r^2) \, .
\ee
We have defined the operators
\be
\Delta X = \sum_i X_{,ii} \  , \qquad
\D(X,Y) = \sum_{i,j} X_{,ij} Y_{,ij} - \Delta X \Delta Y \, .
\ee
with $i,j=1,2,3$ where to leading order, the Gauss-Bonnet term is then 
$\mathcal{L}_{GB} = 8 \D(\Phi,\Psi)$.
For standard Einstein gravity ($V=\alpha=0$), the solution of the
above equations is 
\be
\Phi = \Psi = -U_m \, , \qquad 
\phi = \phi_0 \, ,
\ee
where
\be
U_m = \frac{4 \pi G_0}{c^2} \int d^3 x' \frac{\rho_{m}(\vec
  x',t)}{|\vec x- \vec x'|} \, .
\ee

We will now study solutions which are close
to the post-Newtonian limit of general relativity, and take
\be
\Phi=-U_m+\delta\Phi \, , \qquad
\Psi=-U_m+\delta\Psi \, , \qquad
\phi=\phi_0 +\delta \phi \, ,
\ee
where $\delta \phi$, etc.\ are the leading-order $\alpha$- and $V$-dependent
corrections.

Note that the Laplacian carries a distribution and therefore we have to be
careful with the implementation of the $\D$ operator.
We see that $\delta \phi$ is $\Or(V,\alpha \epsilon^2)$, and so, to leading
order, we have
\be
\Delta \, \delta \phi =V' - 4 \alpha  \xi_1'  \D(U_m,U_m) \, .
\ee
Having calculated $\delta \phi$, we obtain
\bea
\Delta \, \delta \Phi =-V  + 4 \alpha \xi_1' \D(U_m,\delta \phi) 
\\
\Delta \, \delta \Psi=\frac{V}{2} + 2 \alpha \xi_1' \D(U_m, \delta \phi) \, .
\eea

In the case of a spherical distributional source $\rho_m= m \delta^{(3)}(x)$,
\be
U_m= \frac{G_0 m}{c^2 r} \, .
\ee
In accordance to our estimations for $\epsilon$ the solar system
Newtonian potentials are $U_m \lesssim 10^{-5}$, and the velocities satisfy
$v^2 \lesssim U_m$. For planets we have $U_m \lesssim
10^{-7}$ (with the maximum attained by Mercury). 

With the aid of the relation
\be
\D(r^{-n},r^{-m})=\frac{2nm}{n+m+2}\Delta r^{-(n+m+2)} 
\ee
the above expressions evaluate, at leading order, to
\be
\phi = \phi_0 + \frac{r^2 V'}{6} 
- 2\xi_1' \frac{\alpha (G_0 m)^2}{c^4 r^4}
\label{dphi}
\ee
\be
\Phi = -\frac{G_0 m}{c^2 r}\left[1+\frac{8\xi_1'}{3} \alpha V'\right] 
-\frac{r^2 V}{6} 
- \frac{64(\xi_1')^2}{7} \frac{\alpha^2 (G_0 m)^3}{c^6 r^7}
\label{dPhi}
\ee
\be
\Psi = -\frac{G_0 m}{c^2 r}\left[1+ \frac{4\xi_1'}{3} \alpha V' \right] 
+\frac{r^2 V}{12} 
-\frac{32(\xi_1')^2}{7} \frac{\alpha^2 (G_0 m)^3}{c^6 r^7}
\, .
\label{dPsi}
\ee
We find that there are now non-standard corrections to the Newtonian
potential which do not follow the usual parametrised expansion, in
agreement with~\cite{gilles}, but not~\cite{GBPPN} (which uses
different assumptions on the form of the theory).
First of all note that the Gauss-Bonnet coupling $\alpha$ couples to the
running of the dark energy potential $V'$, giving a $1/r$
contribution to the modified Newtonian potential~\eref{dPhi}. We
absorb this into the gravitational coupling,
\be
G= G_0\left[1+\frac{8\xi_1'}{3} \alpha V'\right] \, .
\ee
The corresponding term in~\eref{dPsi} gives a constant
contribution to the effective $\gamma$ PPN parameter.
The $r^2V$ terms in~\eref{dPhi}, \eref{dPsi} are typical of a theory
with a cosmological constant,  whereas the final, $1/r^7$ terms are
the leading pure Gauss-Bonnet correction, which is enhanced at small
distances. If we take the usual expression for the PPN parameter
$\gamma = \Psi/\Phi$, we see that it is $r$ dependent. In using the
Cassini constraint on $\gamma$ we must be careful to calculate the
frequency shift from scratch. 

For the above derivation we have assumed $\delta \phi \ll U_m$, 
which implies $V \ll U_m/r^2$ and $\alpha \ll r^2/U_m$. 
This will hold in the solar system if
\be
V \ll 10^{-36} \m^{-2} \qquad \mbox{and} \qquad
\alpha \ll \left\{ \begin{array}{cc} 10^{23} \m^2 & \mbox{(everywhere)} \\
10^{29} \m^2 & \mbox{(planets only)} \end{array} \right.
\label{valid}
\ee
in geometrised units. Note that strictly speaking there is also a
lower bound on our coupling constants, if the above analysis is to be
valid.  Indeed, if we were to find corrections
of order $\epsilon^2\sim 10^{-14}$, then it would imply that
higher-order corrections from general relativity
were just as important as the ones appearing in~\eref{dPhi}, \eref{dPsi}.

\section{Constraints}
\label{sec:con}

\subsection{Planetary motion}

Deviations from the usual Newtonian potential will affect planetary
motions, which provides a way of bounding them. This idea has been
used to bound dark matter in the solar system~\cite{anderson}, and
also the value of the cosmological constant~\cite{SSlambda}. We will
apply the same arguments to our theory. From the above
gravitational potential~\eref{dPhi}, we obtain the Newtonian acceleration
\begin{equation}
g_{\rm acc}(r) = - c^2 \frac{d\Phi}{dr}
=-\frac{G m}{r^{2}}\left[1-\frac{Vr^{3}}{3r_{g}}
-\frac{64(\alpha\xi_1')^2 r_g^2}{r^{6}}\right]
\equiv -\frac{G m_{{\rm eff}}}{r^{2}}
\label{force}
\end{equation}
where $r_{g} \equiv G m/c^2$ is gravitational radius of the
mass $m$. The above expression gives the effective mass 
$m_{{\rm eff}}$ felt by a body at distance $r$. If the test body is a
planet with semi-major axis $a$, we can use this formula at $r\approx a$. 
Its mean motion $n\equiv\sqrt{G m/a^{3}}$ will then be changed by
$\delta n = (n/2) (\delta m_{\rm eff}/m)$. By evaluating the
statistical errors of the mean motions of the planets, 
$\delta n=-(3n/2)\delta a/a$, we can derive a bound on 
$\delta m_{\rm eff}$ and hence our deviations from general relativity
\begin{equation}
\frac{1}{3}\frac{\delta m_{\rm eff}}{m}=-\frac{Va^{3}}{9r_{g}}
-\frac{64(\alpha\xi_1')^2 r_g^2}{3a^{6}} < \frac{\delta a}{a} \, .
\end{equation}
The values of $a$ for the planets are determined using
Kepler's third law, with a constant sun's mass
$m_{\odot}$. Constraints on $\delta \Phi$ then follow from the errors
$\delta a$, in the measure of $a$. These can be found
in~\cite{pitjeva}, and are also listed in the appendix for convenience.
Given their different $r$-dependence, the two corrections to $\delta
m_{\rm eff}$ are unlikely to cancel. We will therefore bound them
separately, giving constraints on $\alpha$ and $V$.

The strongest bound on the combination $\xi_{1}'\alpha$ comes from 
Mercury, with
\begin{equation}
\left(\frac{\delta a}{a}\right)_{\mercury} 
\lesssim 1.8\times 10^{-12}  \, .\end{equation}
Neglecting the cosmological constant term, and using 
$a\approx 5.8\times 10^{7}\km$ and $r_{g} \approx 1.5\km$, we find
\begin{equation}
|\xi'_{1}\alpha|\lesssim 
\left. \frac{\left(3a^{5} \delta a\right)^{1/2}}{8r_{g}} \right|_{\mercury}
\approx 3.8\times 10^{22}\m^{2} \, .
\label{GBcon1}
\end{equation}
We see that this is within range of validity~\eref{valid} for our perturbative
treatment of gravity.

In cosmology, the density fraction corresponding to the Gauss-Bonnet
term is~\cite{us}
\be
\Omega_{GB} = 4 \xi'_{1}\alpha H \frac{d\phi}{dt} \, .
\ee
If this is to play the role of dark energy in our universe, it needs
to take, along with the contribution of the potential,  a value around $0.7$
at cosmological length scales (and for redshift $z \sim 1$).

If we wish to accurately apply the bound on $\alpha$~\eref{GBcon1} to
cosmological scales, details of the dynamical evolution of $\phi$ will be
required. These will depend on the form of $V$ and the $\xi_i$, and
are expected to involve complex numerical analysis, all of which is
beyond the scope of this work. Here we will instead assume that the
cosmological value of $\phi$ is also $\phi_0$, which, while crude, will
allow us to estimate the significance of the above result.
Given the hierarchy between cosmological and solar system scales 
it is natural to question this assumption but we will make it here, and
discuss it in more detail in the concluding section. 

Making the further, and less controversial, assumption that $d\phi/dt
\approx H$, we obtain a very stringent constraint on $\Omega_{GB}$: 
\begin{equation}
|\Omega_{GB}| \approx  4|\xi'_{1}\alpha|H_{0}^{2}
\lesssim 8.8 \times 10^{-30} \, .
\label{bound1}\end{equation}
Hence we see that solar system constraints on Gauss-Bonnet fraction of
the dark energy are potentially very significant, despite the fact
that the Gauss-bonnet term is quadratic in curvature.

Since we are assuming that all the $\xi_i$ are of the same order, the
above bound also applies to the dark energy fractions arising from the
final three terms in~\eref{Lag}. Clearly there are effective dark
energy models for which the analysis leading to the above
bound~\eref{bound1} does not apply. However any successful model will
require a huge variation of $\xi_1$ between local and cosmological
scales, or a very substantial violation of one of our other
assumptions.

For comparison, we apply similar arguments to obtain a
constraint on the potential. The strongest bound comes from the motion of 
Mars~\cite{SSlambda}, and is 
\be
|V| \lesssim \left. \frac{9 r_g \delta a}{a^4}\right|_{\mars} 
\approx  1.2\times 10^{-40}\m^{-2} \, .
\label{Vcon}
\ee
This suggests $\Omega_V = V/(3H_{0}^{2}) \lesssim 7.3\times 10^{11}$,
which is vastly weaker than the corresponding cosmological constraint 
($\Omega_V \lesssim 1$). Hence planetary orbits tell us little of
significance about dark energy arising from a potential, in sharp
contrast to the situation for Gauss-Bonnet dark energy.

\subsection{Cassini spacecraft}

The most stringent constraint on the PPN parameter $\gamma$ was
obtained from the Cassini spacecraft in 2002 while on its way to
Saturn. The signals between the spacecraft and the earth pass close to
the sun, whose gravitational field produces a time delay. The smallest
value of $r$ on the light ray's path defines the 
impact parameter $b$. A small impact parameter maximises the light
delay. During that year's superior solar conjunction the spacecraft was
$r_{e}=8.43 \AU=1.26 \times 10^{12}\m$ away from the sun, and the
impact parameter dropped as low as $b_{\rm min}=1.6R_{\odot}$. A PPN
analysis of the system produced the strong constraint
\be
\delta \gamma \equiv \gamma-1 = (2.1\pm2.3) \times 10^{-5} \, .
\label{dgamma}
\ee
Given that our theory is not PPN we have to undertake the calculation from
scratch.

The above constraint comes from considering a round trip, in which the
light travels from earth, grazes the sun's `surface', reaches the spacecraft,
and then returns by the same route.
We take the path of the photon to be the straight line between the
earth and the spacecraft, $\vec x = (x,b,0)$ with $x$ varying from
$-x_e$ to $x_\oplus$.  For a round trip (there and back), the
additional time delay for a light ray due to the gravitational field
of the sun is then
\begin{equation}
c \Delta t= 2\int_{-x_e}^{x_\oplus} 
\left[\frac{h_{00}(r)+h_{xx}(r)}{2}\right]dx 
= -2\int_{-x_e}^{x_\oplus} 
\left. (\Phi+\Psi)\right|_{r=\sqrt{x^2+b^2}} dx 
\, .
\end{equation}
For the solution~\eref{dPhi} and~\eref{dPsi}, this evaluates to
\begin{equation} \fl
c \Delta t = 4 r_g \left[1-\frac{2\alpha \xi_1' V'}{3}\right]
\ln \frac{a_\oplus r_e}{4b^2}
+\left[\frac{a_\oplus^3+r_e^3}{3}+b^2(a_\oplus+r_e)\right]\frac{V}{6}
+\frac{1024(\alpha\xi_1')^2 r_g^3}{b^6} \, ,
\label{Dt}
\end{equation}
where we have assumed $x_\oplus \approx a_\oplus \gg b$, and similarly
for the spacecraft.

Rather than directly measure $\Delta t$, the Cassini experiment
actually found the frequency shift in the signal~\cite{cassini}
\begin{equation}
y_{\rm gr} = \frac{d \Delta t}{dt} 
\approx \frac{d \Delta t}{db} \frac{db}{dt} \, .
\end{equation}
The results obtained were
\begin{equation}
y_{\rm gr} = -\frac{10^{-5} \s}{b}\frac{db}{dt}\left(2+\delta\gamma\right)
\, .
\label{cassiniy}
\end{equation}
If gravitation were to be described by the standard PPN formalism, then
$\delta\gamma$ would be the possible deviation of the PPN parameter $\gamma$
from the general relativity value of 1. 

{} From~\eref{Dt} we obtain 
\begin{equation}
y_{\rm gr} =  -\left(2  -\frac{b^2V(a_\oplus+r_e)}{12r_g}
+\frac{1536(\alpha\xi_1')^2 r_g^2}{b^6} - \frac{4\alpha \xi_1' V'}{3}
\right) \frac{4r_g}{cb} \frac{db}{dt} \, .
\label{dgr}
\end{equation}
Requiring that the corrections are within the errors~\eref{dgamma} of
\eref{cassiniy}, implies
\begin{equation}
|\xi'_{1}\alpha|\lesssim \frac{\sqrt{6\, \delta\gamma}}{96} \frac{b^3}{r_g}
\lesssim 1.6 \times 10^{20}\m^{2} \, .
\label{GBcon2}
\end{equation}
This suggests the dark energy bound
\be
|\Omega_{GB}| \lesssim 3.6 \times 10^{-32}\, ,
\ee
although obtaining this bound from solar system data requires major
assumptions about the cosmological behaviour of $\phi$, as we will point out
in section~\ref{sec:dis}.
 
The data obtained by the spacecraft were actually for a range of impact
parameters $b$, but we have just used the most conservative value
$b=b_{\rm min} = 1.6 R_\odot$.  The above constraint is even
stronger than~\eref{GBcon1}, which was obtained for planetary motion. 
This is because the experiment involved smaller $r$, and so the
possible Gauss-Bonnet effects were larger.

Taking the above expression for $y_{\rm gr}$~\eref{dgr} at face value,
we can also constrain the potential to be $|V| \lesssim 10^{-22} \m^2$
and the cross-term $|\alpha \xi_1' V'| \lesssim 10^{-5}$. However
these are of little interest as they are much weaker than the planetary
motion constraints~\eref{GBcon1}, \eref{Vcon}, and also the former is
far outside the range of validity~\eref{valid} of our analysis.

\subsection{A table-top experiment}

Laboratory experiments can also be used to obtain bounds on deviations
from Newton's law. For illustration we will consider the table-top
experiment described in~\cite{hoskins}. It consists of a $60\cm$
copper bar, suspended at its midpoint by a tungsten wire. Two 
$7.3 \kg$ masses are placed on carts far ($105 \cm$) from the bar, and
another mass of $m \approx 43 \g$ is placed near ($5 \cm$) to the side
of bar. Moving the masses to the opposite sides of the bar changes in
the torque felt by it. The experiment measures the torques $N_{105}$ and $-N_5$
produced respectively by the far and near masses. The masses and
distances are chosen so that the two torques roughly cancel. The
ratio $R = N_{105}/N_5$ is then determined, and compared with the
theoretical value.  The deviation from the Newtonian result is 
\be
\delta_R = \frac{R_{\rm expt}}{R_{\rm Newton}}-1 = (1.2 \pm 7) \times 10^{-4}
\label{Rbound}\ .
\ee
In fact, to help reduce errors, additional measurements were
taken. To account for the gravitational field of the carts that the
far masses sit on, the experiment was repeated with only the
carts and a $m' \approx 3\g$ near mass. The measured torque was then
subtracted from the result for the loaded carts.

The Gauss-Bonnet corrections to the Newton
potential~\eref{dPhi} will alter the torques produced by all four
masses, as well as the carts. Furthermore, since $\delta \Phi$ is
non-linear in mass, there will be further corrections coming from cross terms.
The expressions derived in section~\ref{sec:grav} are just for the
gravitational field of a single mass, and so will not fully describe the above
table-top experiment. However, we find that the contribution from the mass
$m$ will dominate the other corrections, and so we can get a good
estimate of the Gauss-Bonnet contribution to the ratio $R$ by just
considering $m$.

The torque experienced by the copper bar, due to a point mass at
$\vec X=(X,Y,Z)$ is
\be
N = \int_{\rm bar} d^3x \, (\vec x \wedge \vec F)_z
= \rho_{\rm Cu} \int_{\rm bar} d^3x \,
\frac{y X -x Y}{r} c^2 \left. 
\frac{d\Phi}{dr}\right|_{r  = |\vec X - \vec x|} \, ,
\ee
where $\rho_{\rm Cu}$ is the bar's density. A full list of parameters
for the experiment is given in table I of~\cite{hoskins}.  The bar's
dimensions are $60 \cm \times 1.5 \cm \times 0.65 \cm$. Working in coordinates
with the origin at the centre of the bar, the mass $m=43.58\g$ is at
$\vec X = (24.42, -4.77, -0.03) \cm$. Treating $m$ as a point mass,
Newtonian gravity implies a torque of 
$N_5 \approx (8.2\cm^2) \, G m \rho_{\rm Cu}$ is produced. The
Gauss-Bonnet correction is
\be \fl
\delta N_5 = \rho_{\rm Cu} \frac{64 G^3 m^3  (\alpha\xi_1')^2}{c^4}
\int_{\rm bar} d^3x \,
 \frac{y X -x Y}{|\vec X - \vec x|^9}
\approx - (0.025\cm^{-4})
\frac{(G m)^3 (\alpha \xi_1')^2 \rho_{\rm Cu}}{c^4}  \ .
\ee
To be consistent with the bound~\eref{Rbound}, we require $\delta N_5/
N_5$ to be within the range of $\delta_R$. This implies
\be
|\alpha \xi_1'| \lesssim (18 \cm^3) \, \frac{c^2 \delta_R^{1/2}}{G m}
\lesssim 1.3 \times 10^{22} \m^2 \, ,
\ee
which is comparable to the planetary
constraint~\eref{GBcon1}. Extrapolating it to cosmological scales gives
\be
|\Omega_{GB}| \lesssim 3.1 \times 10^{-30} \, .
\ee

There are of course many more recent laboratory tests of gravity, and
we expect that stronger constraints can be obtained from
them. Table-top experiments frequently involve multiple gravitational
sources, or gravitational fields which
cannot reasonably be treated as point masses. A more detailed
calculation than the one presented in section~\ref{sec:grav} will then
be required. For example, the gravitational field inside a sphere or
cylinder will not receive corrections of the form~\eref{dPhi}, and so any
experiment involving a test mass moving in such a field requires a
different analysis.

\section{Discussion}
\label{sec:dis}

We have shown that significant constraints on Gauss-Bonnet gravity can be
derived from both solar system measurements and table-top laboratory
experiments (note that further constraints arise when imposing
theoretical constraints like absence of superluminal or ghost modes,
see~\cite{CALC}). The fact that the corrections to Einstein gravity are
second order in curvature suggests they will
automatically be small. However this does not take into account the
fact that the dimensionfull coupling of the Gauss-Bonnet term must be
large if it is to have any hope of producing effective dark energy. 
Additional constraints will come from the perihelion precession of
Mercury, although the linearised analysis we have used is inadequate
to determine this, and higher-order (in $\epsilon$) effects will need to be
calculated.

Performing an extrapolation of our results to cosmological scales
suggests that the density fraction $\Omega_{GB}$ will be far too small
to explain the accelerated expansion of our universe.
This agrees with the conclusions of~\cite{gilles}. Hence if
Gauss-Bonnet gravity is to be a viable dark energy candidate, one needs
to find a loophole in the above arguments. This is not too
difficult, and we will now turn to this question.

In particular, we have assumed no spatial or temporal evolution of the
field $\phi$ between cosmological and solar system scales, even
though the supernova measurements correspond to a higher redshift and
a far different typical distance scale. A varying
$\phi$ would of course imply that different values of $\xi_i$, and their
derivatives, would be perceived by supernovas and the planets. It is
interesting to note that the size of the bound we have
found~\eref{bound1} is of order the square of the ratio of the solar
system and the cosmological horizon scales,
$s=(1\AU \, H_0)^2\sim 10^{-30}$. Therefore one could reasonably argue 
that the small number appearing in~\eref{bound1} could in fact be
due to the hierarchy scale, $s$, rather than a very stringent constraint
on $\Omega_{GB}$. This could perhaps be concretely realised with something
similar to the chameleon effect~\cite{chameleon} giving some constraint 
on the running of the quintessence theory. One other
possibility is that the baryons (which make up the solar system) and
dark matter (which is dominant at cosmological scales) have different
couplings to $\phi$~\cite{CQ}. Again, this would alter the relation between
local and cosmological constraints.

Alternatively, it may be that our assumptions on the form of the
theory should be changed. The scalar field could be coupled directly
to the Einstein-Hilbert term, as in Brans-Dicke gravity. Additionally,
the couplings $\xi_i$ and their derivatives could be of different
orders. The same could be true of the potential. In particular, if
$\phi$ were to have a significant mass, this would suppress the
quadratic curvature effects, as they operate via the scalar
field. This would be similar to the situation in scalar-tensor gravity
with a potential, where the strong constraints on the theory can be
avoided by giving the scalar a large mass (which, however, would inhibit
acceleration).

Finally, the behaviour of the scalar field could be radically
different. We took it to be $\Or(\epsilon)$, like the metric
perturbations. However since our constraints are on the metric, and
not $\phi$, this need not be true. Furthermore, since the theory is
quadratic, there may well be alternative solutions of the field
equations, and not just the one we studied.

Hence to obtain a viable Gauss-Bonnet dark energy model, which is compatible
with solar system constraints, at least one of the above assumptions must be
broken. For many of the above ideas the higher-order scalar kinetic terms will
play a significant role. This then opens up the possibility that the
higher-gravity corrections will cancel each other, further weakening the
constraints. We hope to address some of these issues in the near future. 

\ack
CC thanks Martin Bucher, Gilles Esposito-Farese and Lorenzo Sorbo for
discussions. SCD thanks the Netherlands Organisation for Scientific
Research (NWO) for financial support.

\section*{Appendix}

For the benefit of readers without an astronomical background, we list
relevant solar system parameters. The values for $\delta a$ come
from table 4 of~\cite{pitjeva}. We take the
Hubble constant to be $H_0 = 70 \km\s^{-1}{\Mpc}^{-1}$.

\begin{minipage}{3in}
\begin{tabular}{|l|}
\hline
$R_{\odot}=6.96\times10^{8}\m$\tabularnewline
$r_{g}^{\odot} \equiv G m_{\odot}/c^{2}=1477\m$\tabularnewline
$H_{0}/c=7.566\times10^{-27}\m^{-1}$\tabularnewline
$1 \AU\equiv a_\oplus =149597870691\m$\tabularnewline
\hline
\end{tabular}
\vspace{0.2in}

\begin{tabular}{|l|}
\hline 
$G=6.6742\times10^{-11}\m^{3}\s^{-2}\kg^{-1}$\tabularnewline
$c=299792458\m\s^{-1}$\tabularnewline
$m_{\odot}=1.989\times10^{30}\kg$\tabularnewline
\hline 
\end{tabular}
\end{minipage}
\begin{minipage}{3in}
\begin{tabular}{|l|c|c|}
\hline 
name &
$a$ ($10^{9}\m$) &
$\delta a$ (m) \tabularnewline
\hline 
Mercury &
57.9 &
0.105 \tabularnewline
Venus &
108 &
0.329 \tabularnewline
Earth &
149 &
0.146 \tabularnewline
Mars &
228 &
0.657 \tabularnewline
Jupiter &
778 &
639 \tabularnewline
Saturn &
1433 &
$4.22\times10^{3}$ \tabularnewline
Uranus &
2872 &
$3.85\times10^{4}$ \tabularnewline
Neptune &
4495 &
$4.79\times10^{5}$ \tabularnewline
Pluto &
5870 &
$3.46\times10^{6}$ \tabularnewline
\hline
\end{tabular}
\end{minipage}

\end{document}